\documentstyle[12pt,bloisnew]{article}
\def\mum{\,\mu{\rm m}}
\begin{document}
\heading{
COSMOLOGICAL EVOLUTION OF THE SUBMILLIMETRE LUMINOSITY OF 
HIGH-REDSHIFT RADIO GALAXIES
}

\author{D.H. Hughes$^{1}$, J.S. Dunlop$^{1}$, E.N. Archibald$^{1}$,
S. Rawlings$^{2}$, S.A. Eales$^{3}$} 
{$^{1}$ Institute for Astronomy, University of Edinburgh, U.K.}
{$^{2}$ Astrophysics, Nuclear \& Astrophysics Lab., University of Oxford, U.K.}
{$^{3}$ Dept. of Physics \& Astronomy, University of Cardiff, U.K.}

\begin{bloisabstract}
A systematic survey measuring the submillimetre continuum luminosity
in radio galaxies between redshifts $z \sim 0.1 - 5$ 
is currently in progress.  The first results
from observations with the bolometer array SCUBA on the JCMT suggest a
trend of increasing submillimetre luminosity with redshift out to 
$z \simeq 4$.  Assuming
the continuum emission at 850$\mum$ is dominated by thermal radiation
from dust heated by young, massive stars, the straightforward
interpretation of these data implies that the host galaxies of
powerful radio sources, presumably ellipticals or their progenitors,
exhibit increased star formation activity with increasing
redshift. However a severe bias may distort the true picture since
only a subset of 30 galaxies have been observed, which represent the
most luminous radio sources ($P_{151 \rm MHz} > 10^{27.5} \rm\,W Hz^{-1}
sr^{-1}$), whilst the complete sample covers $\sim 4$ decades in radio power
($\rm 10^{25} - 10^{29} \,W Hz^{-1} sr^{-1}$).
This on-going observational programme continues to improve
coverage of the $P-z$ plane in an attempt to remove this
potential bias towards higher AGN activity 
and hence understand more fully the relationship between
high-z AGN, their level of starformation, and the evolutionary status of
their host galaxies.
\end{bloisabstract}

\section{Introduction}
The last 12 months have seen submillimetre astronomy make a significant
impact on cosmological studies of galaxy formation and evolution.
This is primarily due to the technological development of bolometer
arrays now operating on the worlds largest submillimetre and
millimetre telescopes.  There currently exist two distinct,
feasible and complementary observational programmes at submillimetre
wavelengths which address important questions regarding the
evolutionary status of high-redshift galaxies. These can be summarised as:

(1) Blank field submillimetre 
surveys at 850$\mum$, covering a range of depths 
(with  3$\sigma$ detection limits between 1.5 and 8\,mJy)  and areas
(between 0.002 - 0.1 sq. degrees)  
\cite{Nature} \cite{Barger} \cite{Smail} \cite{Eales}, 
with the aim of determining the cosmological evolution of the
starburst galaxy population.

(2) Pointed observations of known high-redshift AGN, including radio galaxies
\cite{HDR}, radio-loud and radio-quiet quasars \cite{Omont}, designed to
determine the epoch of elliptical galaxy formation, if indeed a unique
epoch exists, and the cosmological evolution of dust mass 
in massive elliptical galaxies.

Both observational programmes are motivated by several lines of
evidence, outlined below, 
that suggest an era of massive star-formation exists at 
high-redshift.

\begin{figure}[t]
\begin{center}
\setlength{\unitlength}{1mm}
\begin{picture}(200,100)
\includegraphics{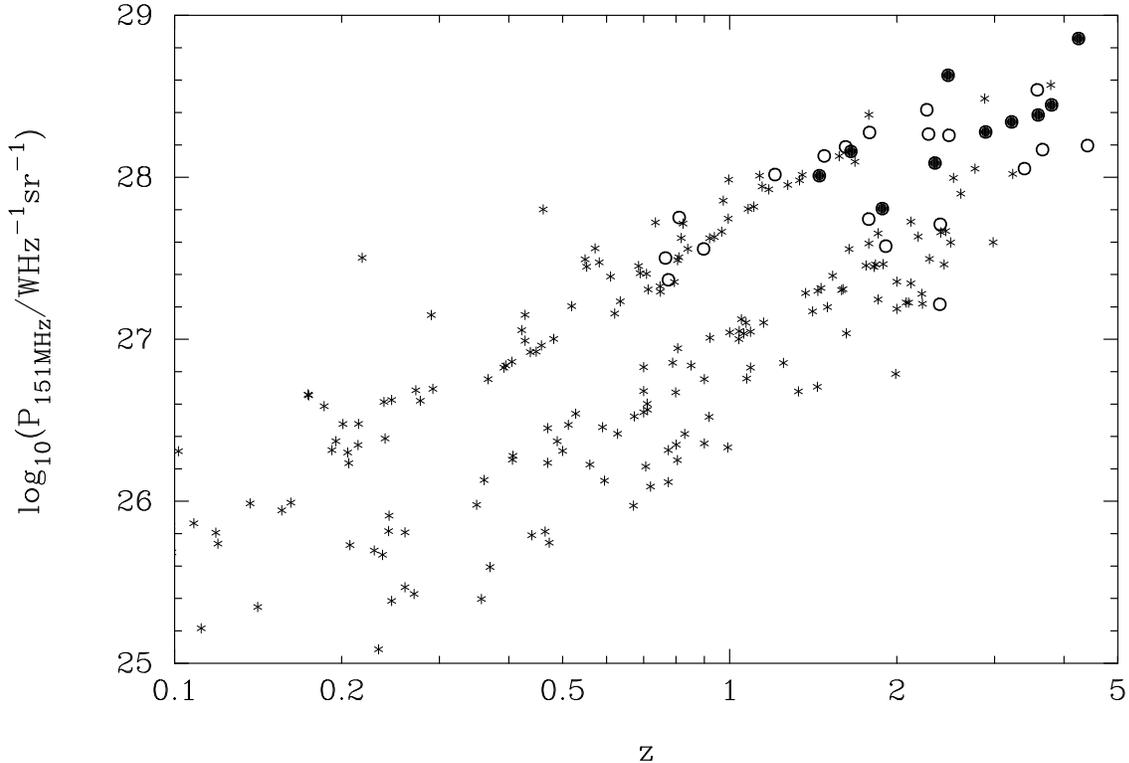}
\end{picture}
\caption[]{\scriptsize 
The radio-luminosity:redshift ($P - z$) plane of high-z radio-loud galaxies.
The entire SCUBA sample of $> 200$ radio galaxies (stars), selected from
progressively deeper radio surveys (3C, 6C, 7C, LBDS), covers a large
range in radio power and redshift.  The circles represent those
targets already observed with SCUBA, for which the open and solid
symbols indicate non-detections and detections ($> 3\sigma$) at 850$\mum$ 
respectively. The subset of completed SCUBA observations have been
confined to a region of parameter space defined by log$(P_{\rm 151 MHz}/{\rm
WHz^{-1}sr^{-1}} ) > 27.0$, $z > 0.7$.
Using sub-mm observations of the complete sample we can 
quantify the contribution of an AGN to the rest-frame FIR   
luminosity, and trace the evolution of gas mass and star-formation rate
as function of redshift and radio luminosity.
}
\end{center}
\end{figure}

(i) First, strong evidence exists for a redshift cutoff in the number density
of radio-loud sources, indicating that AGN activity peaked at
redshifts $z = 2-3$ \cite{DP}. Second, deep infrared K-band imaging has
demonstrated that in the low-z universe ($z < 1$) luminous radio
sources, and all optical quasars (both radio-loud and radio-quiet)
more luminous than $M_{V} = -24$, reside in giant elliptical
galaxies with K-band luminosities $> 2 L_{\star}$ and hence stellar
masses $> 5 \times 10^{11}  M_{\odot}$ \cite{Taylor}. If it can be
assumed that the host galaxies of powerful high-redshift AGN
are also massive ellipticals and are fully assembled at $z \sim 2$,
then a sustained star-formation rate (SFR)
$> 250 M_{\odot}/\rm yr$ is required during the 
previous 2--3 Gyrs of their evolution.

(ii) The elliptical hosts of weak radio sources at $z \sim 1.5$
show absorption line spectra consistent with stellar populations
having a formation epoch $z_{f} > 4$ \cite{Dunlop98}.

(iii) The clustered populations of Lyman-limit galaxies at $z \sim 3$ 
show modest average SFRs of $1-5 \,h^{-2} M_{\odot}/\rm yr$ 
\footnote[2]{$h = 100\,{\rm kms^{-1}} Mpc^{-1}$}.  However these observed
SFRs, calculated from rest-frame UV luminosities may significantly
underestimate the true SFRs by factors of $\sim 2-15$ due to the efficient
absorption of UV radiation by dust \cite{Heckman}, \cite{Pettini}.

(iv) The traditional interpretation of number counts in the deepest
optical surveys suggests that the star formation density peaked at
$z \sim 1-1.5$, a factor of 10 greater than the present-day value,
and declined at higher redshift. However there is now conflicting
evidence, based on deep submillimetre surveys \cite{Nature} and corrections
for incompleteness at high-z in the optical surveys \cite{Pascarelle},
which suggests that the rate of star formation may remain high ($0.2 \,h
M_{\odot} \rm yr^{-1} Mpc^{-3}$) at $z \sim 3$.

Thus regardless of whether one is considering the most luminous AGN,
weak radio sources, Lyman-break galaxies or starburst galaxies
identified in submillimetre surveys it appears that active star
formation may have proceeded in the most massive systems at a rate $\gg
100 M_{\odot}/\rm yr$ in the early universe.

\section{The evolution of star formation rate in massive elliptical galaxies}

Whilst dust is a disadvantage in the optical/UV, the presence of dust
heated by young, massive stars means that at early epochs elliptical
galaxies can be expected to be not only bright in the rest-frame FIR
\cite{Mazzei}, but also to show a significant evolution in their FIR
luminosities from their formation to the present epoch. At the highest
redshifts this FIR spectral peak is shifted into the submillimetre regime
making observations at 850$\mum$ a powerful method to quantify the
starformation history of the massive elliptical galaxies that host
radio-loud high-z AGN. 

The spectral shape of a typical starburst galaxy in the
submillimetre/FIR means that a large negative K-correction is produced
at submillimetre wavelengths. At 850$\mum$ this effect is so great
that a galaxy of fixed FIR luminosity has a approximately 
constant flux density over the redshift range
$z \sim 1-10$ \cite{HDR}. Consequently the
bolometer array SCUBA \cite{SCUBA}, operating on the 15-m JCMT, is now
able to detect in a few hours a galaxy undergoing a burst of star
formation similar in intensity to that in the low-z ULIRG Arp220 at
any redshift out to $z \sim 10$ \cite{HD}.

In this paper we expand on our earlier work \cite {HDR}
and present the first results from an extensive series 
of SCUBA observations aimed at measuring the rest-frame FIR properties 
of a large sample of high-z, steep-spectrum, radio-loud AGN covering 
the radio-luminosity:redshift ($P-z$) plane
($25.0 < {\rm log}\,P_{151 \rm MHz}/\rm W Hz^{-1}sr^{-1}  < 28.9$, 
$0.1 <  z < 5$), as shown in figure\,1. 

Submillimetre measurements of radio-loud sources selected from various
surveys (6C, 7C, 8C, LBDS) lead to an estimate of the dust masses, gas
masses, FIR luminosities and SFRs in high-z elliptical galaxies, as
well as the dependence of these physical properties on 
redshift (for a fixed radio luminosity) or AGN
luminosity (for a given redshift bin).
 
The preliminary SCUBA observations of 30 radio galaxies, whilst 
concentrating on the highest radio-power sources
at all redshifts $z > 0.7$, have also endeavoured to provide a
sub-sample with a narrower range (less than a factor 3) in radio
luminosity and a similarly broad range in redshift ($z \sim 1 - 5$).
Whether one
considers the weighted-mean $850\mum$ flux densities, binned in redshift,
of all observed radio galaxies spanning a range of $\sim 45$ in radio power, 
or the subset drawn from a significantly smaller range of radio power 
($28.0 < {\rm log}\, P_{151 \rm MHz}/\rm W Hz^{-1} sr^{-1}  < 28.5$), 
both samples show a trend of increasing submillimetre flux
density with increasing redshift (figure\,2), extending to $z > 3$.  

The data obtained to date thus suggest that the dust enshrouded star
formation in the massive ellipticals which host radio-loud AGN
increases monotonically with redshift out to $z > 3$. This result
would clearly be consistent with a primary formation epoch of massive
ellipticals at $z \simeq 4$. However, we caution that at present
sub-mm luminosity and radio luminosity remain correlated in our sample
(although less strongly correlated than $S_{850\mum}$ and
redshift). Further sub-mm observations should allow us to remove this
bias and exploit high-redshift radio-galaxies as unbiased tracers of
the evolution of SFR in massive ellipticals in general.

\begin{figure}[t]
\begin{center}
\setlength{\unitlength}{1mm}
\begin{picture}(200,100)
\includegraphics{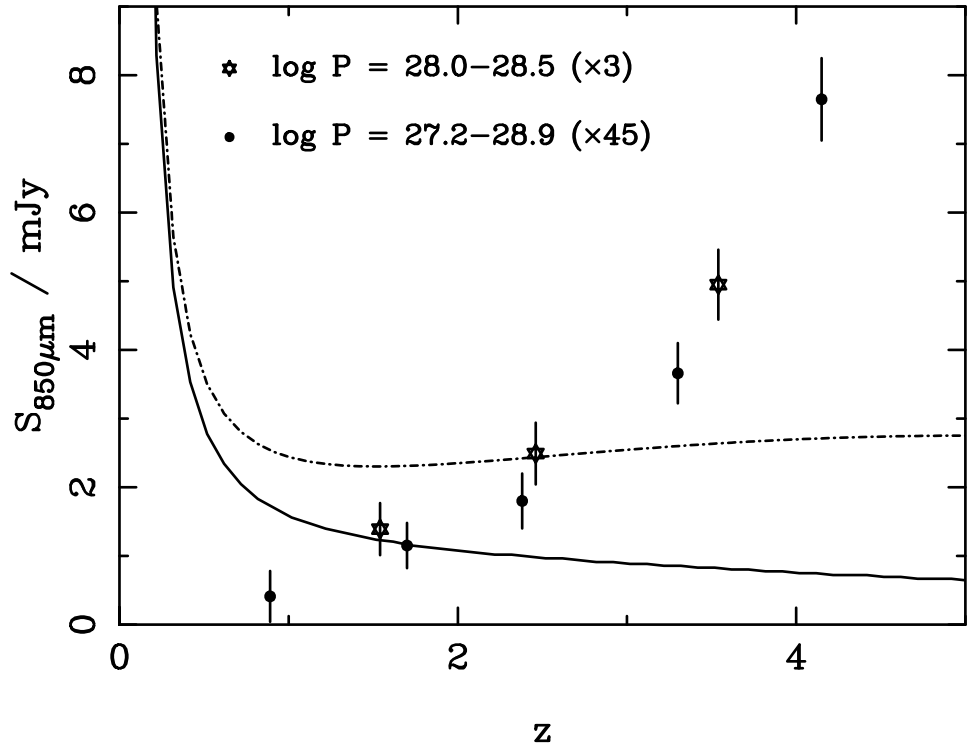}
\end{picture}
\caption[]{\scriptsize Submillimetre flux density vs. redshift for
radio-loud AGN.  The weighted-mean $850\mum$ flux densities of the
entire sample of radio galaxies (solid circles) observed to date and
the subset confined to a narrow band of AGN power, (log\,$P_{151 \rm
MHz}/\rm W Hz^{-1} sr^{-1} = 28.0 - 28.5$, shown as open stars), are binned
in redshift. In both cases the overall trend is for submillimetre flux
density to increase with redshift and hence the result may be
independent of AGN power. Given the strong negative K-correction at
850$\mum$ which results in a galaxy of fixed rest-frame FIR luminosity
having approximately constant flux density between redshifts $z
\sim 1-10$, this increase in
submillimetre flux density suggests an increase in the star-formation
rate in elliptical galaxies or their progenitors with redshift. 
The dotted-dashed and solid curves represent
the 850$\mum$ flux-density for the redshifted spectrum of Arp220
assuming $\Omega_{0} = 1.0$ and 0.1 respectively. 
}
\end{center}
\end{figure}

 

%
\vfill
\end{document}